\documentclass[12pt,preprint]{aastex}
\usepackage[dvips]{color}
\usepackage{graphicx}

\def\msol{\hbox{\kern 0.20em $M_\odot$}}
\def\lsol{\hbox{\kern 0.20em $L_\odot$}}
\def\rsol{\hbox{\kern 0.20em $R_\odot$}}
\def\sr{\hbox{\kern 0.20em sr}}
\def\srmu{\hbox{\kern 0.20em sr$^{-1}$}}
 
\def\g{\hbox{\kern 0.20em g}}
\def\gmu{\hbox{\kern 0.20em g$^{-1}$}}
\def\kg{\hbox{\kern 0.20em kg}}
\def\pc{\hbox{\kern 0.20em pc}}
 
\def\mum{\hbox{\kern 0.20em $\mu$m}}
\def\mumd{\hbox{\kern 0.20em $\mu$m$^{-2}$}}
\def\cm{\hbox{\kern 0.20em cm}}
\def\m{\hbox{\kern 0.20em m}}
\def\km{\hbox{\kern 0.20em km}}
\def\nm{\hbox{\kern 0.20em nm}}
 
\def\s{\hbox{\kern 0.20em s}}
\def\h{\hbox{\kern 0.20em h}}
\def\sec{\hbox{\kern 0.20em sec}}
\def\min{\hbox {\kern 0.20em min}}
\def\smu{\hbox{\kern 0.20em s$^{-1}$}}
\def\smd{\hbox{\kern 0.20em s$^{-2}$}}
\def\an{\hbox{\kern 0.20em an}}
\def\anmu{\hbox{\kern 0.20em an$^{-1}$}}
\def\deg{\hbox{\kern 0.20em $^{\rm o}$}}
\def\yr{\hbox{\kern 0.20em yr}}
\def\yrmu{\hbox{\kern 0.20em yr$^{-1}$}}
\def\Myr{\hbox{\kern 0.20em Myr}}
\def\Mymu{\hbox{\kern 0.20em Myr$^{-1}$}}
\def\K{\hbox{\kern 0.20em K}}
\def\pcmu{\hbox{\kern 0.20em pc$^{-1}$}}
\def\pcmd{\hbox{\kern 0.20em pc$^{-2}$}}
\def\pcmt{\hbox{\kern 0.20em pc$^{-3}$}}
\def\kms{\hbox{\kern 0.20em km\kern 0.20em s$^{-1}$}}
\def\kmpd{\hbox{\kern 0.20em km$^{2}$}}
\def\kpc{\hbox{\kern 0.20em kpc}}
\def\cms{\hbox{\kern 0.20em cm\kern 0.20em s$^{-1}$}}
\def\erg{\hbox{\kern 0.20em erg}}
\def\ergs{\hbox{\kern 0.20em erg}}
\def\cmpd{\hbox{\kern 0.20em cm$^2$}}
\def\cmmd{\hbox{\kern 0.20em cm$^{-2}$}}
\def\cmms{\hbox{\kern 0.20em cm$^{-6}$}}
\def\cmpt{\hbox{\kern 0.20em cm$^3$}}
\def\cmmt{\hbox{\kern 0.20em cm$^{-3}$}}
\def\mpd{\hbox{\kern 0.20em m$^2$}}
\def\mmd{\hbox{\kern 0.20em m$^{-2}$}}
\def\mpt{\hbox{\kern 0.20em m$^3$}}
\def\mmt{\hbox{\kern 0.20em m$^{-3}$}}
\def\mujy{\hbox{\kern 0.20em $\mu$Jy}}
\def\mjy{\hbox{\kern 0.20em mJy}}
\def\Mj{\hbox{\kern 0.20em MJy}}
\def\jy{\hbox{\kern 0.20em Jy}}
\def\ghz{\hbox{\kern 0.20em GHz}}
\def\srmd{\hbox{\kern 0.20em sr$^{-1}$}}
\def \kms{km~$\rm{s}^{-1}$}

\def \mum{$\mu$m}
\def\asec{$^{\prime\prime}$}

\def\Mjysr{\hbox{\kern 0.20em MJy~sr$^{-1}$}}
\def\jybkms{\hbox{\kern 0.20em Jy~bm$^{-1}$~km~s$^{-1}$}}
\def\mjb{\hbox{\kern 0.2em mJy bm$^{-1}$}}

\def\G{\hbox{\kern 0.20em G}}

\def\h13cop{\hbox{H$^{13}$CO$^{+}$}}

\def\S+{\hbox{S{\small II}}}

\shorttitle{ISMEVOL}
\shortauthors{Young, Bendo \& Lucero}

\begin{document}

\newcommand{\jfourteen}{\hbox{$J=14\rightarrow 13$}}
 \title{Star Formation in an Unexpected Place: Early-type Galaxies}

\author{Lisa M.\ Young\altaffilmark{1},
George J.\ Bendo\altaffilmark{2}, and
Danielle M.\ Lucero\altaffilmark{1}
}

\altaffiltext{1}{New Mexico Tech, Socorro, NM, USA}
\altaffiltext{2}{Imperial College, London, UK}

\begin{abstract}
Many early-type galaxies are detected at 24 to 160\mum\ but the emission is
usually dominated by an AGN or heating from the evolved stellar population.
Here we present MIPS observations of a sample of elliptical and lenticular
galaxies which are rich in cold molecular gas, and we investigate how much
of the MIR to FIR emission could be due to star formation activity.  The
24\mum\ images show a rich variety of structures, including nuclear point
sources, rings, disks, and smooth extended emission, and comparisons to
matched-resolution CO and radio continuum images suggest that the bulk of
the 24\mum\ emission could be traced to star formation.  The star formation
efficiencies are comparable to those found in normal spirals.  Some future
directions for progress are also mentioned.
\end{abstract}


\lefthead{Young et al.}

\righthead{Star Formation in Early-type Galaxies}

\section{Introduction}

In recent years, UV and optical photometry and spectroscopy of nearby
elliptical galaxies has suggested that these galaxies, which have a
reputation for being old, red, and dead, 
may not be quite as dead as previously assumed.
Some 15\% to 30\% of local ellipticals seem to be experiencing small amounts 
of present 
day star formation activity (Schawinski et al.\ 2007a, 2007b; Kaviraj et al.\ 2007).
The star formation is not intense enough to cause serious problems
for the galaxies' morphological classification, as it only amounts to a few
percent of the total stellar mass.  
However, this current day disk growth inside spheroidal galaxies may be a
faint remnant of a process which was more vigorous in the
past and may have played a role in establishing the spectrum of galaxy
morphologies we observe today.

Star formation of course requires cold gas, so interpreting the UV and optical
data in terms of star formation activity has important implications both for
the early-type galaxies and for a general understanding of the star
formation process.  It is not obvious that star formation should ``work" the
same way inside spheroidal galaxies as it does inside disks, with the same
efficiency or the same dependence on the gas surface density.
For example, it has been hypothesized that even if there is a molecular disk
inside an elliptical or lenticular galaxy, the disk would probably be
stabilized by the galaxy's steep gravitational potential (e.g.\ Kennicutt
1989; Okuda et al.\ 2005; Kawata et al.\ 2007).  Thus it is of interest to
probe the relationships between molecular gas and star formation activity in
early-type galaxies.

It is not as straightforward to measure star formation rates in early-type
galaxies as it is in spirals, however.  Optical line emission is common in
ellipticals (Shields 1991; Goudfrooij et al.\ 1994; Sarzi et al.\ 2006),
but usually its widespread
and/or filamentary distribution and its line ratios indicate that it is more closely
related to AGN activity or to cooling from hot gas than to star formation.
Far-IR and cm-wave radio continuum emission are also 
commonly used as tracers of star formation activity in gas-rich spirals, but 
the mid-IR and far-IR emission in ellipticals is usually attributed to AGN activity
and/or the evolved
stellar population rather than to star formation (Temi et al.\ 2007, 2008).
Here we investigate some evidence for and against star formation activity in a
sample of elliptical and lenticular galaxies that have unusually large molecular gas
contents (Table 1).  We make use of matched-resolution images of the molecular gas
distribution, the cm-wave radio continuum and the 24\mum\ intensity.

\def\htoo{$\rm H_2$}
\def\solmass{$\rm M_{\sun}$}
\def\solum{$\rm L_{\sun}$}

\begin{deluxetable}{ccccccc}
\tablewidth{0pt}
\tablecaption{Sample Galaxies -- Basic Properties}
\tablehead{
\colhead{Name} & \colhead{Type} & \colhead{Dist} & \colhead{log L$_B$} &
\colhead{log L$_{60\mu m}$} & \colhead{M(\htoo)} & \colhead{environment} \\
\colhead{} & \colhead{} & \colhead{(Mpc)} & \colhead{(\solum)} &
\colhead{(L$_{B,\odot}$)} &
\colhead{$10^8$ \solmass} & \colhead{} \\
}
\startdata
UGC 1503 & E & 67 & 10.2 & 9.4 & 25 & field \\
NGC 0807 & E & 63 & 10.5 & 9.2 & 19 & field \\
NGC 2320 & E & 77 & 10.8 & 9.2 & 43 & group dominant \\
NGC 3032 & S0 & 21 & 9.7 & 9.0 & 5.0 & field \\
NGC 3656 & I0 & 42 & 10.2 & 9.7 & 64 & merger remnant \\
NGC 4459 & S0 & 16 & 10.2 & 8.7 & 1.6 & Virgo Cluster \\
NGC 4476 & E/S0 & 17 & 9.5 & 8.3 & 1.5 & Virgo Cluster \\
NGC 4526 & S0 & 16 & 10.5 & 9.2 & 5.7 & Virgo Cluster \\
NGC 5666 & Ec/S? & 34 & 9.8 & 9.4 & 7.8 & field \\
\enddata
\end{deluxetable}

\section{Molecular Gas in Early-type Galaxies}

Early surveys for molecular gas in early-type galaxies were strongly biased
towards FIR-bright targets, but more recent work is not biased in this way and still
finds significant molecular gas contents.
Welch \& Sage (2003)
found a surprisingly high CO detection rate of $78\%$ in a
volume-limited sample of nearby field lenticular galaxies, and
Sage et al.\ (2007) detected CO emission in $33\%$ of a similar sample of
field ellipticals.  Combes et al.\ (2007) also detected CO emission in $28\%$ of
the early-type galaxies in the SAURON survey (de Zeeuw et
al.~2002), a representative sample which uniformly fills an optical
magnitude -- apparent axis ratio space.  Thus, the CO detection rates 
may be high enough to support the UV-inferred incidence of star formation activity. 
The cold gas masses are highly variable in these detections, 
with $M_{\rm gas}/L_B$ in the range $10^{-1}$
to $10^{-3}$ and lower.  

Since we are interested in morphology as a means of distinguishing the origin of the
radio and IR emission, we have selected for this project elliptical and lenticular
galaxies with maps resolving their molecular gas distribution.  If the 24\mum\
emission arises in star formation activity, we expect it to trace the molecular gas.
If the 24\mum\ emission is related to an AGN we expect it to be a point source, and
if it is circumstellar dust it should trace the stellar distribution.  Thus, we also
required the targets to have molecular gas which is extended enough that it should be
resolved in the 24\mum\ images.  
The molecular gas maps are published by Young (2002, 2005) and
Young, Bureau, \& Cappellari (2008); 1.4 GHz (20cm) radio continuum maps are
available from the VLA FIRST survey and from Lucero \& Young (2007).

\section{Comparison samples: CO-poor early-type galaxies}

A handful of elliptical and lenticular galaxies were observed as part of the SINGS
survey and their morphologies are discussed by Bendo et al.\ (2007).  With 
a couple of exceptions, such as 
NGC~1316 (which contains an extended arc of emission) and NGC~5866
(which contains an edge-on disk), these
early-type galaxies tended to have highly compact, symmetric nuclear pointlike 
or nearly-point
sources at 24\mum.  At least three of these are detected in CO emission but CO maps are
not yet available.


MIPS observations of elliptical galaxies have also been published by Temi et al.\
(2007, 2008) and Kaneda et al.\ (2007).  Of the 19 galaxies discussed by Temi et al.\
(2008), 13 have been searched for CO emission and none have been detected.  In these
works the 24\mum\ emission from the ellipticals follows the near-IR surface
brightness profiles very closely, so the 24\mum\ surface brightness is
close to a $r^{1/4}$ profile with the same effective radius as in the $K$-band.  In
addition, Temi et al.\ (2007) have shown that the 24\mum\ emission globally tracks
the optical luminosity in elliptical galaxies as there is a tight linear correlation
between the 24\mum\ flux density and the $B$-band flux density.   This 24\mum\
emission is interpreted to be circumstellar dust from the mass loss of post main
sequence stars.
Thus, in the
majority of the elliptical galaxies that are not CO-rich the 24\mum\ emission 
seems to either follow the stellar photospheric emission or a nuclear source.

\section{Results: 24\mum\ morphologies}


Simple model fits are made in order to provide some parametrization of the 24
\mum\ morphologies.  We first constructed an empirical PSF from archival
observations of 3C 273, 3C 279, and BL Lac.  Point sources,
exponential disks, rings, and de Vaucouleurs $r^{1/4}$ profiles (and
combinations of these) are convolved with the PSF and fit to the images, just
as was done for the Sombrero Galaxy by Bendo et al.\ (2006).  None of the
CO-rich early-type galaxies are pure point sources at 24\mum; all are resolved, 
but are still significantly less extended than the stellar distributions in the
NIR and optical.  

The 24\mum\ intensity is a function of both the dust surface density and the
illuminating radiation field.
Possible dust heating sources include 
the post-main sequence stellar population, star formation regions, and AGN.
Thus, the 24\mum\ emission by itself may not necessarily indicate the presence of
star formation.
However, star formation is expected to be accompanied by spatially resolved cm-wave
radio continuum emission (Condon 1992), whereas dust heated by the 
radiation from evolved stars would not be. 
Therefore the comparisons with the distribution of the molecular gas (the raw
material for star formation) and the radio continuum 
provide constraints on the origin of the 24\mum\ emission.

\begin{figure}
\centerline{
\includegraphics[scale=1.0]{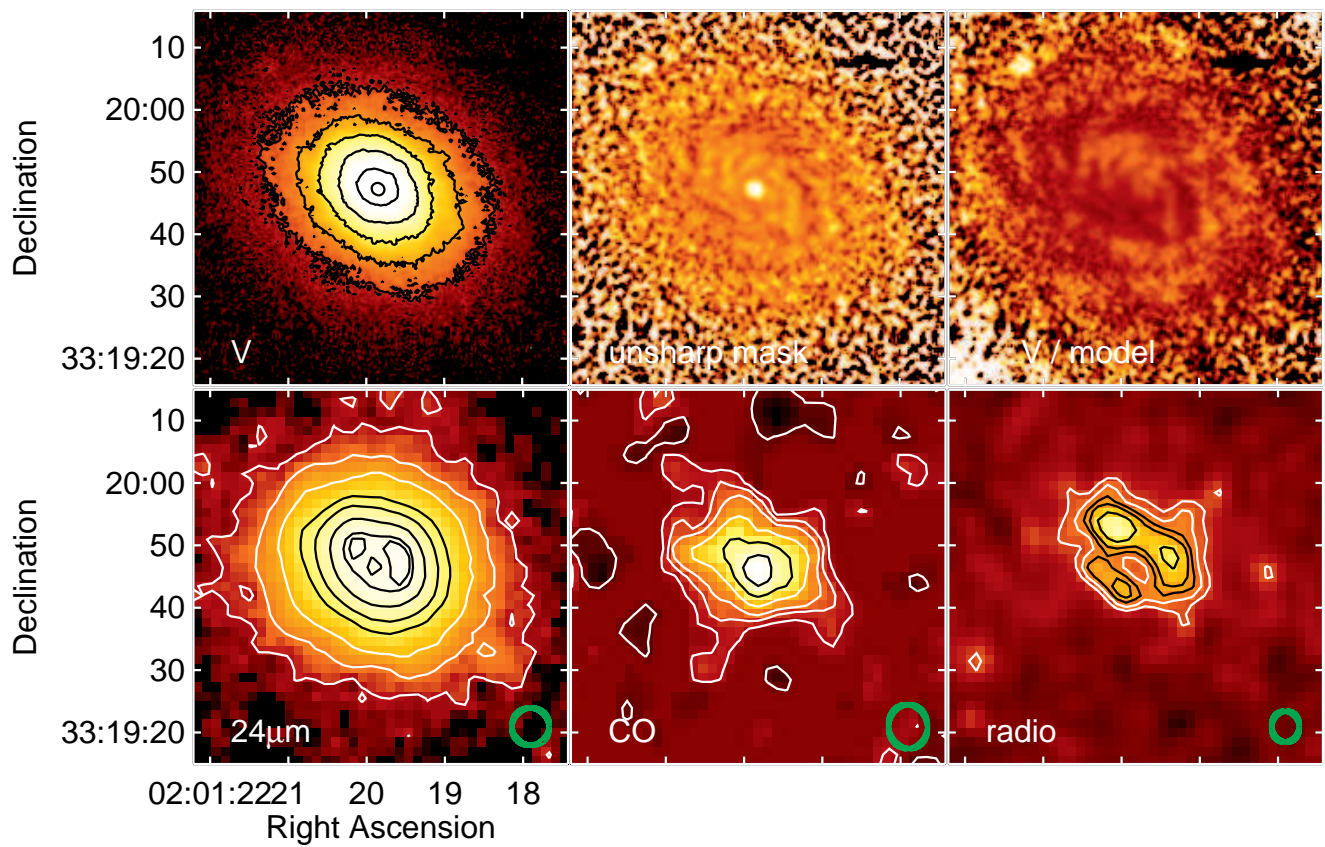}
}
\caption{\label{u1503} \small Optical, IR, CO and radio continuum morphology of
UGC~1503.  Optical contours are spaced by a factor of two.  The top row, center
panel is an unsharp-masked $V$ image and the left panel is the ratio of the
$V$ to a smooth Multi-Gaussian Expansion model (Cappellari 2002).  Contour levels
in the 24\mum\ image are 2, 5, 10, 20, 30, 50, 70, and 75 percent of the peak
(6.1 \Mjysr).  Contour levels in the CO image are $-10$, 10, 20, 30, 50, 70,
and 90 percent of the peak (6.3 \jybkms).  Contours in the 1.4 GHz radio
continuum image are 0.09, 0.12, 0.15, 0.18, and 0.24 \mjb.  Green circles show the
spatial resolution.
}
\end{figure}

\begin{figure}
\centerline{
\includegraphics[scale=1.0,bb=0 0 540 320,clip]{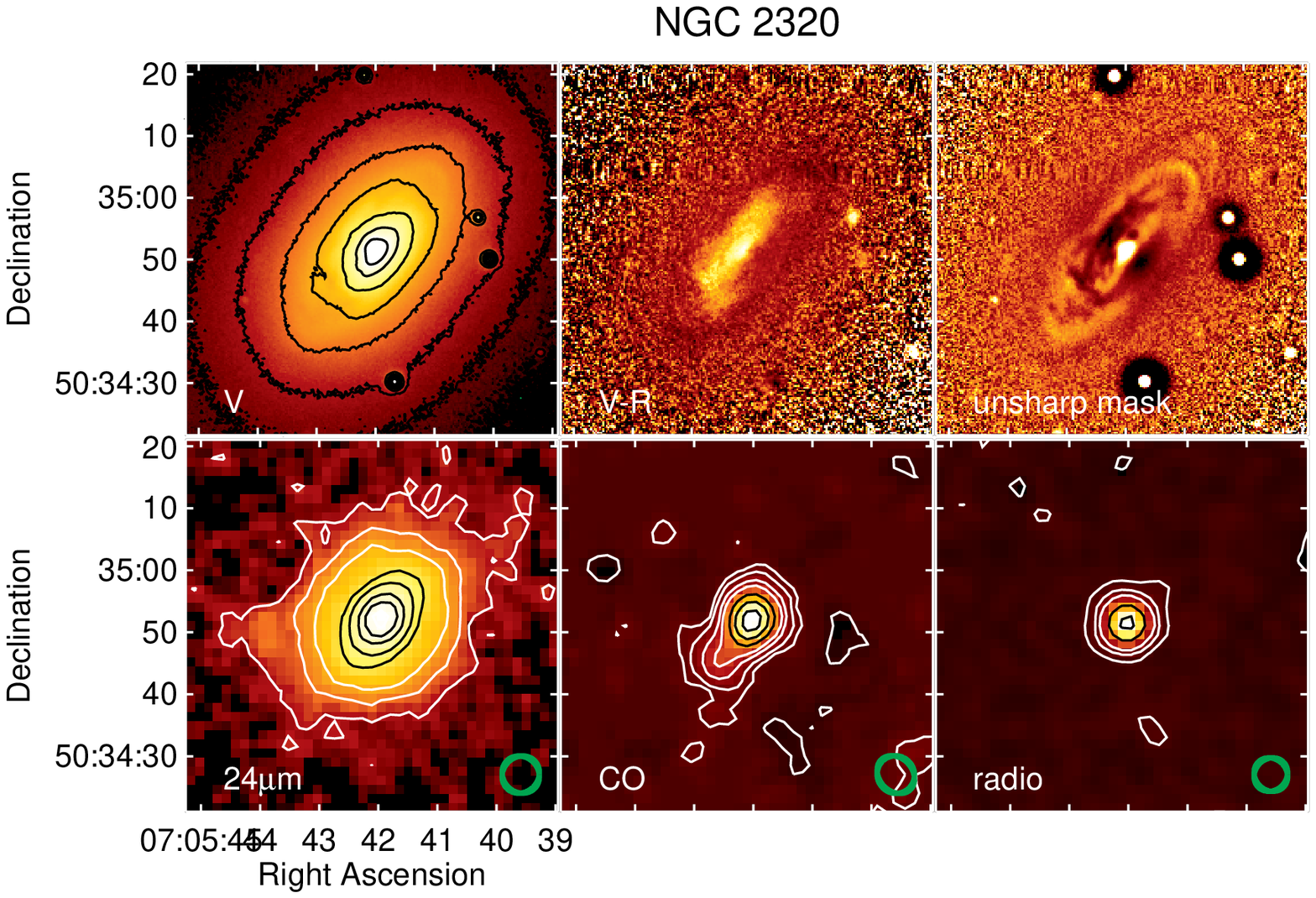}
}
\caption{\label{n2320} \small Optical, IR, CO and radio continuum morphology of
NGC~2320.  Optical contours are spaced by a factor of two.  Contours in the
24\mum\ image are 2, 5, 10, 20, 30, 50, 70, and 90 percent of the peak (5.6
\Mjysr).  Contours in the CO image are $-5$, 5, 10, 20, 30, 50, 70, and 90
percent of the peak (29.1 \jybkms).  Contours in the 1.4 GHz radio continuum
image are $-3$, 3, 10, 20, 50, and 90 percent of the peak (13.1 \mjb).
Green circles show the beam sizes (FWHM).
}
\end{figure}

Figures \ref{u1503} and \ref{n2320} show some of the variety of
morphologies in these CO-rich early-type galaxies.  
For example, UGC 1503 shows a regularly rotating molecular gas (and dust) disk of
diameter 30\asec\ = 9.7 kpc, with a similarly sized ring of radio continuum 
emission.  The 24\mum\ emission also shows a ring with a central dip, and
both radio continuum and 24\mum\ have morphologies consistent with
star formation in the molecular gas.  Most of the other cases in the sample
also have closely matching radio, 24\mum\ and CO morphologies, and they are
interpreted in a similar manner.

\begin{figure}
\centerline{
\includegraphics[scale=0.6]{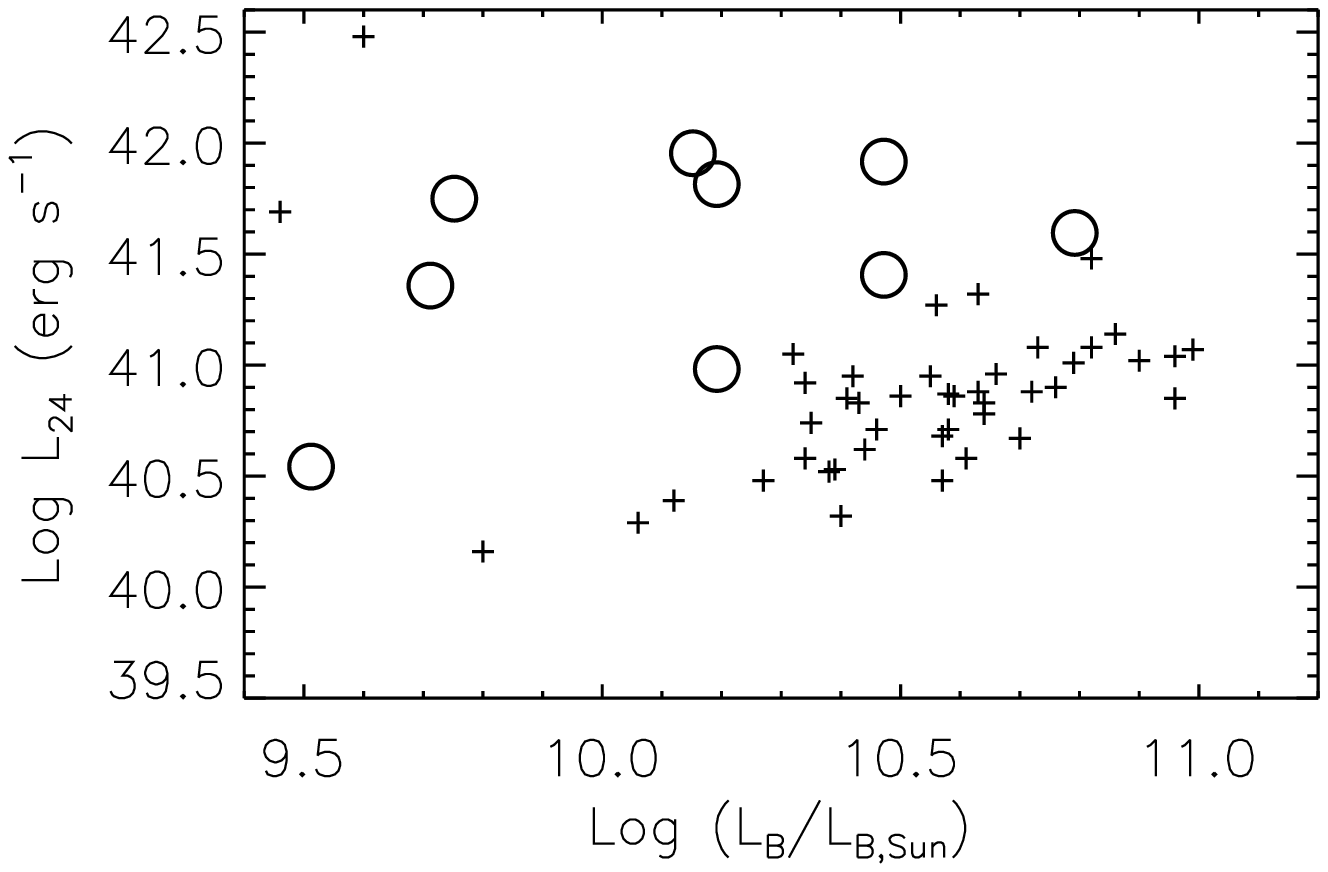}
\includegraphics[scale=0.6,bb=135 252 522 540,clip]{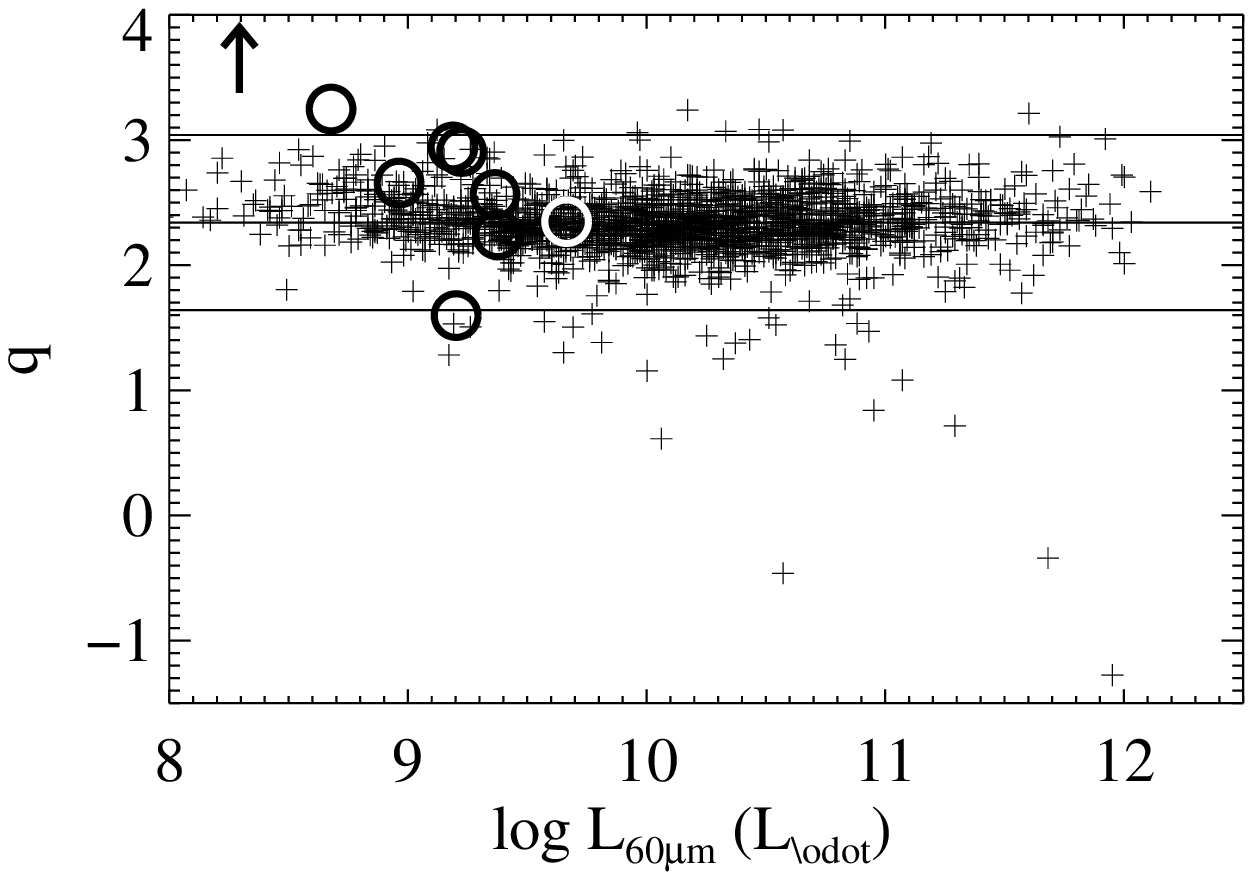}
}
\caption{\label{temifig} \small (Left) Comparison of the 24\mum\ luminosity and optical
luminosity for the CO-rich early-type galaxies (circles) and the generally
CO-poor ellipticals of Temi, Brighenti, \& Mathews 2008 (crosses).
(Right) FIR-to-radio flux ratio $q$.  The circles and the arrow
are the CO-rich
early-type galaxies studied here and crosses are the data of Yun, Reddy, \&
Condon (2001).  As defined in Yun et al., the flux ratio $q$ has a mean value of
2.34 for star-forming galaxies (mostly spirals); the lines at $q=3.04$ and
$q=1.64$ indicate the IR excess and radio excess boundaries, respectively, at 
roughly
$2.7\sigma$ from the mean.  
}
\end{figure}

In the case of NGC~2320 it is not at all clear whether the 24\mum\ emission is
driven by star formation.
The galaxy's radio continuum emission is more likely due to AGN activity than to
star formation (Young 2005).
The 24\mum\ emission is well fit by a smooth $r^{1/4}$ model with 
half-light semimajor and semiminor axes of 4\asec $\times$ 2\asec, but this is five times
smaller than the effective radius of the stellar distribution.  
This modestly extended 24\mum\ emission certainly could be driven by star
formation in the inner part of the molecular gas disk, 
but star formation activity is not required.

\section{IR luminosities and star formation efficiencies}

Figure \ref{temifig} compares the 24\mum\ and optical luminosities of CO-rich
and CO-poor early-type galaxies.  
The ones rich in molecular gas are typically a factor of 10 more luminous at
24\mum, for a given optical luminosity, than would be expected if their 24\mum\
emission were entirely circumstellar in origin.
Our morphological analysis has also shown that the 24\mum\ morphology is not that
of point sources and is a
much better match to the molecular gas distributions than to the stellar light
profiles.
In addition, Figure \ref{temifig} shows that the FIR/radio continuum flux
density ratios are (with the exception of NGC~2320) consistent with star
formation activity.  
Based on this evidence we believe it is reasonable to attribute the bulk of the
24\mum\ emission in the CO-rich early-type galaxies to star formation.
Under the assumption that the same is true for the FIR emission in the IRAS 60\mum\
and 100\mum\ bands, Combes et al. (2007) found that the implied star formation 
efficiencies in the SAURON early-type galaxies 
are similar to those in normal spirals (Combes et al.\ 2007).
The IRAS data gave no useful morphological information, though, so our analysis
provides the validation for this star formation assumption.

\section{Summary and open questions}

For the majority of the CO-rich early-type galaxies the close
agreements between CO, 24\mum, and radio continuum morphologies suggest that
the bulk of the 24\mum\ emission should be attributed to star formation
activity.  Radio and FIR flux density ratios are consistent with this
interpretation, as are the increased $L_{24\mu m}/L_B$ ratios of the CO-rich
over the CO-poor early-type galaxies.
The CO, radio, and MIPS data are thus roughly
consistent with the UV results implying star formation activity in a few tens
of percent of the nearby early-type galaxies.
The necessary raw material is
present, more or less, and the molecular gas is often being turned into
stars.
Detailed comparisons with the UV morphologies should be made as well.

Future work should also make more quantitative tests of theoretical 
and phenomenological models of the star formation process.  For example, 
with a model of the gravitational potential (from the stellar
distribution and a mass-to-light ratio) one can test whether the Toomre-type 
local gravitational instability is consistent with the locations of star
formation activity, as discussed by Kawata et al.\ (2007).  The gas-rich
early-type galaxies could also provide useful perspective on the workings of
the Kennicutt-Schmidt relationship between the star formation rate and the gas
surface density.  If these models have truly captured some underlying physics
of the star formation process they ought to work in the ellipticals and
lenticulars as well as in the spirals.

Where the molecular gas came from to begin with is, in general, still an open question.
In this regard comparisons of the kinematics of the gas and
stars will be useful, as shown in Young et al.\ (2008), as will correlations
between the gas content and the stellar structural and dynamical indicators 
of the galaxy's history. 
The gas-to-dust ratios might also help to distinguish whether the molecular gas was
acquired from a low metallicity source such as a dwarf galaxy or from a progenitor 
of roughly solar metallicity.

\acknowledgements

This work is based on observations made with the {\it Spitzer Space
Telescope}, which is operated by the Jet Propulsion Laboratory (JPL),
California Institute of Technology under NASA contract 1407.
Support for this work was provided by NASA and through JPL Contract 1277572.

\end{document}